\documentclass[usenatbib]{mn2e}
\usepackage{graphicx,amssymb,amsmath,rotating}
\title[X-ray heating of circumstellar discs]{The effects of X-ray photoionization and heating on the structure of circumstellar discs}

\author[R.D.~Alexander, C.J.~Clarke \& J.E.~Pringle]
  {R.D.~Alexander\thanks{email: rda@ast.cam.ac.uk},
  C.J.~Clarke
  and J.E.~Pringle\\
Institute of Astronomy, Madingley Road, Cambridge, CB3 0HA}
\begin{document}

\pagerange{\pageref{firstpage}--\pageref{lastpage}} \pubyear{2004}

\maketitle

\label{firstpage}
\begin{abstract}
We present the results of a theoretical study investigating the effects of photoionization and heating by X-rays on discs around low-mass stars.  In particular we address the question of whether or not X-rays can drive a disc wind.  First, we construct a 1-dimensional ``quasi-hydrostatic'' model, which solves for the vertical structure introduced by X-ray heating.  We consider uniform X-ray illumination of the disc, but the X-ray fluxes required to heat the disc significantly are much greater than those seen by recent observations.  When the model is extended to consider heating from a central X-ray source we find that the 1-dimensional model is only valid very close to the star.  We extend our analysis to consider a simple 2-dimensional model, treating the disc as a two-layered structure and solving for its density profile self-consistently.  For T Tauri stars we are able to set a crude upper limit on the mass-loss rate that can be driven by X-ray photoevaporation, with a value of $\simeq 10^{-13}$g cm$^{-2}$ s$^{-1}$.  Our model is designed to maximise this value, and most likely over-estimates it significantly.  However we still find a mass-loss rate which is less than that found in studies of ultraviolet photoevaporation.  We conclude that in the presence of a significant UV field, X-ray driven disc winds are unlikely to play a significant role in the evolution of discs around low-mass stars.
\end{abstract}



\section{Introduction}
The existence of discs around young low-mass stars is now well established and remains an important area of study, as disc evolution provides strong constraints on theories of both star and planet formation.  The majority of young stars are observed to have circumstellar discs at ages $\sim 10^6$yr \citep*{strom89,kh95,haisch01}.  These discs are optically thick in the optical and infrared, and typically have masses of a few percent of a solar mass \citep*{beckwith90,ec03}.  However these discs are not observed in most systems at ages of $\sim 10^7$yr: generally only low-mass ``debris discs'' remain by this point (e.g.~\citealt*{ms97,wyatt03}).  Thus disc lifetimes are expected to be $\sim 10^6$yr.  

However few objects are observed to have properties intermediate between those of disc-bearing, classical T Tauri stars and discless, weak-lined T Tauri stars \citep*{kh95,act99}.  Thus very few objects are ``caught in the act'' of dispersing their discs, and therefore the dispersal timescale must be short, $\sim 10^5$yr \citep{sp95,ww96,duvert00}.  The common mechanisms invoked as means of disc dispersal, such as magnetospheric clearing or viscous evolution, fail to satisfy this two-time-scale constraint \citep{act99}.  Instead they produce power-law declines in disc properties \citep{hcga98}, thus always producing dispersal time-scales of the same order as the disc lifetime.

One promising avenue of investigation, however, has been using photoevaporative heating as a means of driving disc dispersal.  Models of the flows driven from steady discs by ultraviolet (UV) photoevaporation were first constructed by \citet{holl94}, and have since been improved upon by several subsequent studies (\citealt*{yw96,ry97,kyr98}; see also the review by \citealt*{hollppiv}).  These models are characterised by mass loss outside some critical radius (usually referred to as the ``gravitational radius''), beyond which ionized material is heated to above the escape temperature.  The mass loss rate declines sharply with radius (\citealt{holl94} find an $R^{-5/2}$ scaling beyond the gravitational radius), and so most of the mass loss occurs at or very near to the gravitational radius.  Recently \citet{font04} have performed detailed hydrodynamic modelling of this process, and whilst they find the mass-loss profile around the gravitational radius to be modified somewhat, their results are in broad agreement with those from the earlier models.  

Initially these models failed to reproduce observed disc lifetimes, as the low mass-loss rates meant that the disc was dispersed much too slowly.  However more recent studies \citep*{cc01,mjh03,acp03} have coupled the photoevaporative mass loss with viscous evolution, and found that a number of observed disc properties {\it can} be satisfied by these models.  In particular, the ``UV switch'' model of \citet{cc01} was able to reproduce the two-time-scale behaviour described, and so these models provide an attractive area of study.  However a number of problems still exist.  The models fail to satisfy observational constraints on the sub-mm emission associated with the outer parts of the disc at late times \citep{cc01}, and detailed SED modelling has not yet been performed.  Further, and more importantly, the details of the Lyman continuum emission from the central star, emission required to drive the photoevaporation, are poorly understood.  Even the magnitude of the emission is poorly constrained \citep{gahm79,ia87}.  In a previous paper we have shown that emission from the accretion shock is unlikely to drive disc photoevaporation \citep*{columns}, and it is not known if stellar chromospheres can provide the requisite flux of ionizing photons.  Further, given the problems associated with observing at wavelengths immediately shortward of the Lyman limit, it is not clear when, or if, this issue will be resolved.  

A logical step beyond UV photoionization models is to look at the effect of X-ray ionization and heating on disc evolution.  X-rays are obviously capable of ionizing disc material and, unlike in the UV case, the X-ray emission from young stellar objects and T Tauri stars (hereafter TTS) has been well studied (e.g.~reviews by \citealt{neuhauser97,fm99} and references within).  X-rays are emitted from both magnetic reconnection events near to the star and also from protostellar outflows/jets, in both cases with luminosities potentially large enough to influence disc evolution.  Further, recent observations of young stellar clusters containing massive stars (of type earlier than O5) have found diffuse X-ray emission throughout the clusters \citep{townsley03}.  The effect of this external X-ray flux on disc structure and evolution may also warrant investigation.

Previous studies of X-ray/disc interactions (e.g.~\citealt*{gni97,ig99,ftb02}) have focused mainly on using X-rays as a means of sustaining the low levels of ionization required to drive the magnetorotational instability \citep{bh91}, or studied the effect of heating to low temperatures ($\le 1000$K) on the observed spectrum \citep{gn01,gh04}.  Here we make a preliminary study of X-ray heating as a means of driving a disc wind.  The structure of the paper is as follows.  In Section \ref{sec:xrays} we review relevant observations and X-ray physics, and then set up our basic disc model.  In Section \ref{sec:uniform} we use this 1-dimensional (1-D) model to study a disc subject to uniform X-ray illumination, and then apply the model to illumination from a central source in Section \ref{sec:central}.  In this case the model is of limited validity, and so in Section \ref{sec:2d} we extend it to a simple 2-dimensional (2-D) model.  In Section \ref{sec:dis} we discuss the caveats and limitations that apply to the model, and in Section \ref{sec:conc} we summarize our conclusions.


\section{Disc winds and X-ray heating}\label{sec:xrays}
\subsection{Background}\label{sec:bg}
Previous studies of photoevaporatively driven disc winds have focused on the case of UV photoionization, where the disc is heated by photons with energies of 13.6eV.  This has been studied in detail by a number of authors (e.g.~\citealt{holl94,ry97,font04}; see also the review chapter by \citealt{hollppiv}).  The 2-D radiative transfer scheme of \citet{holl94} finds that a central UV source produces a hot ionized atmosphere above the surface of the isothermal disc.  They find that the diffuse field (ionizing photons produced by recombinations of H{\sc i} atoms in the atmosphere) dominates over the direct field from the central source.  Beyond a gravitational radius, the thermal energy of the particles in the atmosphere becomes greater than their gravitational potential energy, and so beyond this gravitational radius ionized particles escape from the disc: this is photoevaporative mass-loss.

By comparison to the UV case the case of X-ray heating is more complex, due to the more complicated nature of the interaction of X-rays with matter.  The physical processes involved in X-ray absorption by diffuse gas are described in detail by \citet{kk83} (see also the reviews by \citealt{fm99} and \citealt{x_ppiv}).  Here we merely summarise the salient points.

The X-ray emission around Young Stellar Objects (YSOs) and TTS is well studied, and is due to three distinct mechanisms.  First, and best-studied, is the X-ray emission from the central star and its immediate surroundings.  This is thought to arise from magnetic field reconnection events in the stellar corona and magnetosphere, and typically produces an X-ray spectrum consistent with thermal bremsstrahlung (e.g.~\citealt{neuhauser97,fm99}).  This emission peaks at at around 2keV, and YSOs and TTS typically exhibit steady X-ray luminosities of around $10^{28}$--$10^{30}$erg s$^{-1}$.  They also exhibit rapid variability and occasional ``flaring'', when large changes in X-ray luminosity, sometimes several orders of magnitude, occur on timescales of order hours (e.g.~\citealt*{ikt01}).

The second source of X-ray emission around young stars is from the jets and shocks observed in protostellar outflows.  Such sources are visible at other wavelengths as Herbig-Haro (HH) objects, and produce X-ray emission with similar properties (spectrum, luminosity etc.) to that produced by the corona \citep{fm99}.  However these jets and shocks are far from the central star and disc, at distances typically tens to hundreds of AU.  This geometric dilution of the energy, combined with probable extinction due to absorbing material between the HH shock and the disc, means that this X-ray emission is unlikely to have any significant effect on the disc.  Therefore we do not include this emission in our modelling.

The third mechanism is a very recent discovery.  \citet{townsley03} observed X-ray emission throughout M17 and the Rosette Nebula, and attributed it to thermal bremsstrahlung emission from the diffuse ionized ISM.  They observed integrated luminosities of order $10^{33}$erg s$^{-1}$ from areas on the sky of order $\sim 50$pc$^2$.  Little variation was observed across the spatial extent of the clusters, and the observed spectrum was slightly softer than that seen from magnetospheric X-rays (peaking at 0.8--1.5keV).  However this emission has only been in observed in clusters containing very massive stars.  Indeed the lack of any similar emisssion in the Orion Nebula Cluster \citep{feigelson03} indicates that stars of spectral type O5 or earlier may be required in order to produce this emission in young stellar clusters.  It follows that this cannot be a universal mechanism for disc dispersal, as low-mass star forming regions such as Taurus-Auriga contain no massive stars.  It may be relevant in some cases, however, and is also much simpler to analyse than the case of emission from a central source.

When soft X-ray emission is incident on the gas in a disc incident X-ray photons are usually absorbed by heavy elements, such as O, C or Fe.  A single photoelectron is produced, which then collides with neutral material in the gas, generating typically 30 ``secondary'' electrons.  Heavy elements can also undergo the Auger effect, in which the excited ion produced by the initial photoabsorption undergoes 2-electron decay, and direct (collisional) charge exchange between neutral hydrogen or helium and ionized heavy species can also occur.  A key difference bewteen the X-ray and UV cases is the details of the incident radiation field.  Whilst in the case of UV ionization the recombination (diffuse) field dominates, the probability of an X-ray photon being emitted by a recombination is small \citep{gni97}.  Thus only the direct field is important in X-ray heating, and the heating is local to the initial X-ray absorption.  Consequently care must be taken with regard to the geometry of the problem.  Of the 3 cases discussed above, we neglect the emission from the HH outflow.  First, we consider the geometrically simpler case of uniform illumination from a diffuse plasma, of the type observed by \citet{townsley03}.  Having studied the effects of uniform illumination we then discuss the more promising, but also more complicated, case of X-rays generated by magnetic activity near to the surface of the central object.

Throughout the following calculations we adopt an incident X-ray spectrum consistent with that of $10^7$K optically thin bremsstrahlung, with a peak at $\simeq 0.7$keV.  These energies are typical of both the diffuse and coronal emission observed (see discussion above).  In order to ensure that the models consider only the effect of the X-rays and not the ``tail'' of the spectrum at lower energies, we subject the bremsstrahlung spectrum to an exponential cutoff at energies $E < 0.1$keV.

\subsection{Basic disc model}\label{sec:basic_model}
As a first iteration we make use of a simple hydrostatic disc model.  We set up a steady disc, and then consider the effects on the structure of X-ray heating.  Initially we treat the radiation as being vertically incident on the disc, and so we make the approximation that the disc can be treated as a series of concentric, non-interacting annuli.  Here we define our steady disc model.

The surface density, $\Sigma(R)$, of the steady disc is given by
\begin{equation}\label{eq:surf_den}
\Sigma(R)=\frac{M_d}{2\pi R_s R}\exp(-R/R_s)
\end{equation}
This represents a power-law decline in surface density surface density (c.f.~\citealt{beckwith90,bell97}) proportional to $1/R$, tapered exponentially at around a scaling radius $R_s$.  We note however that our results do not depend strongly on the exact form of $\Sigma(R)$.  We also note that this is similar to the form of theoretically derived time-dependent solutions for the surface density (e.g.~\citealt{lbp74,hcga98,cc01}).  We adopt a typical disc mass of $M_d=0.01$$M_{\odot}$.  We adopt a scaling radius of $R_s=10$AU initially, and later consider the effect of varying this parameter.

At a given radius $R$ the vertical structure is that of an isothermal disc (see \citealt{pringle81}).  The general equation of hydrostatic equilibrium perpendicular to the plane of the disc is
\begin{equation}
\frac{1}{\rho} \frac{\partial P}{\partial z} = \frac{\partial}{\partial z}\left(\frac{GM_*}{\sqrt{R^2+z^2}}\right)
\end{equation}
However for small vertical displacements $z$ above or below the midplane we can approximate the gravitational potential to be harmonic, and so the equation of hydrostatic equilibrium simplifies to
\begin{equation}\label{eq:hydroeq}
\frac{1}{\rho} \frac{\partial P}{\partial z} = -\frac{GM_*z}{R^3} \quad  \quad (\text{for } z\ll R)
\end{equation}
As shown in \citet{pringle81} we can solve this to find the isothermal vertical density structure, which is a Gaussian distribution in $z$:
\begin{equation}\label{eq:iso_den}
\rho(z) = \rho_0(R) \exp\left(-\frac{z^2}{2H^2(R)}\right)
\end{equation}
Here the density $\rho_0(R)$ is given by
\begin{equation}
\rho_0(R)=\frac{\Sigma(R)}{\sqrt{2\pi}H(R)}
\end{equation}
where the scale height $H(R)$ is defined to be
\begin{equation}
H(R)=\left(\frac{c_s^2(R) R^3}{GM_*}\right)^{1/2}
\end{equation}
and $c_s$ is the local isothermal sound speed,
\begin{equation}
c_s^2(R)=\frac{P}{\rho} = \frac{k T(R)}{\mu m_{\mathrm H}}
\end{equation}
Thus choosing the disc temperature to be a function of $R$ determines the disc structure uniquely.  The behaviour of the disc temperature with radius is not completely understood, and so power-laws are usually adopted.  Simple reprocessing of stellar radiation by a thin disc results in a $T \propto R^{-3/4}$ scaling \citep{as86}, and accretion-powered ``self-luminosity'' results in the same scaling relationship \citep{lbp74}.  However such a relationship does not fit observed data, and tends to under-predict the flux observed at long wavelengths \citep{kh87}.  Much work has been done on this (e.g.~see discussions in \citealt{kh87,hcga98}), and the consensus is that a $T \propto R^{-1/2}$ scaling law is both consistent with observational data and theoretically justifiable (for example the ``flared reprocessing disc'' proposed by \citealt{kh87}).  Therefore, again following \citet{cc01}, we adopt a power-law scaling of the form 
\begin{equation}
T(R)=\left(\frac{R}{R_0}\right)^{-1/2} T_0
\end{equation}
with the normalisation condition $T_0=100$K at $R_0=1$AU \citep{beckwith90}.  We adopt a fiducial stellar mass of $M_* = 1M_{\odot}$ throughout.


\section{Uniform illumination}\label{sec:uniform}
\subsection{Modelling X-ray heating}\label{sec:1dmodel}
In order to study what we expect to happen to such a disc when it is subjected to, and subsequently heated by, ionizing radiation, we have first made use of the {\sc cloudy} photoionization code \citep{cloudy}.  Given a fixed, static density structure this code solves the equations of thermal and ionization equilibium to find a unique solution.  Only a density structure, incident flux spectrum and chemical compsition must be specified as input parameters\footnote{We assume solar abundances throughout.}.  When run, {\sc cloudy} returns temperature and ionization profiles for the heated region.  It also highlights the main heating and cooling processes. 

Initially, we treat the disc as a series of concentric, non-interacting annuli heated vertically from above.  This reduces the problem to a series of independent 1-D problems.  We take the vertical density profile of the steady disc and use this as an input to the {\sc cloudy} code.  If the sound crossing timescale, given by $t_s \sim c_s/H(R)$, is much shorter than the thermal equilibrium timescale then we can assume that the disc will ``relax'' to a non-isothermal vertical structure much faster than the heating processes change.  Thus we can take the temperature profile obtained from {\sc cloudy} and solve the equation of hydrostatic equilibrium in the vertical direction to find a new, updated density profile.  This is in turn used as an input to {\sc cloudy}, and so we iterate towards a self-consistent solution.  In practice the sound crossing timescale is always found to be at least 10 times less than the thermal equilibrium timescale, and so this approximation is valid.

When we solve the equation of hydrostatic equilibrium (Equation \ref{eq:hydroeq}) in the non-isothermal case, we find that the particle number density $n(z)$ is given by
\begin{equation}\label{eq:den_iter}
n(z)=n_R \frac{T_R}{T(z)}\exp\left(- \frac{m_{\mathrm H}}{k} \frac{GM_*}{R^3} \int_{z_R}^z \frac{\mu(z') z'}{T(z')} dz'\right)
\end{equation}
for $z>z_R$.  In this equation $T(z)$ is the temperature, $n(z)$ the particle number density, $\mu(z)$ the mean molecular weight, and the subscripted values $T_R$ and $n_R$ refer to the values of these functions evaluated at some reference point $z_R$, which is introduced as a boundary condition.  This solution is presented in full in Appendix \ref{sec:app_a}.

In practice, as can be seen from Equation \ref{eq:den_iter}, the choice of initial density structure has a strong effect on this process, and not all initial structures converge towards a self-consistent solution.  In order to do this we adopt a ``two-component'' initial profile, consisting of a cold central core (that of the unperturbed, isothermal disc) and a hot ``atmosphere'' in the region where we expect the disc to be heated.  The core has the standard isothermal structure above (Equation \ref{eq:iso_den}).  The heated atmosphere has a similar structure, but this time with a scale height
\begin{equation}
H_{\text {hot}}=\left(\frac{c_s^2 R^3}{GM_*}\right)^{1/2} = \left(\frac{k T_{\text {hot}} R^3}{\mu m_{\mathrm H} GM_*}\right)^{1/2}
\end{equation}
For the regions of interest, temperatures of 1000--10,000K are typical, and so we adopt a value of $T_{\text {hot}}=3000$K for this initial guess.  We note however that as long as the procedure converges the results do not depend on this value.  For the inital profile we define a reference height $z_T$ at which temperature changes from the midplane temperature to $T_{\text {hot}}=3000$K.  We maintain pressure equilibrium at this point, and so the inital density profile shows a contact discontinuity at this point.  The criterion for determining $z_T$ is discussed below.

Our model is limited by the fact that the {\sc cloudy} code is only valid for temperatures $\ge 3000$K, and so we cannot solve explicitly for the structure all the way down to the disc midplane.  Consequently we are forced to extrapolate from the height at which the temperature falls to 3000K down to the point at which it reaches the midplane temperature ($z_R$ in Equation \ref{eq:den_iter}).  We extrapolate linearly, keeping the gradient of $T(z)$ constant from the stopping point of the code (3000K) down to the midplane.  In doing this we neglect the existence of an extended column of ``warm'' material ($\sim 1000$K).  Other studies, while exploring somewhat different parameter spaces, have found that such an extended column probably does result when circumstellar discs are heated by X-rays \citep{gn01,gh04}, which may modify the resulting mass-loss rates somewhat.

\begin{figure}
        \resizebox{\hsize}{!}{
        \begin{turn}{270}
        \includegraphics{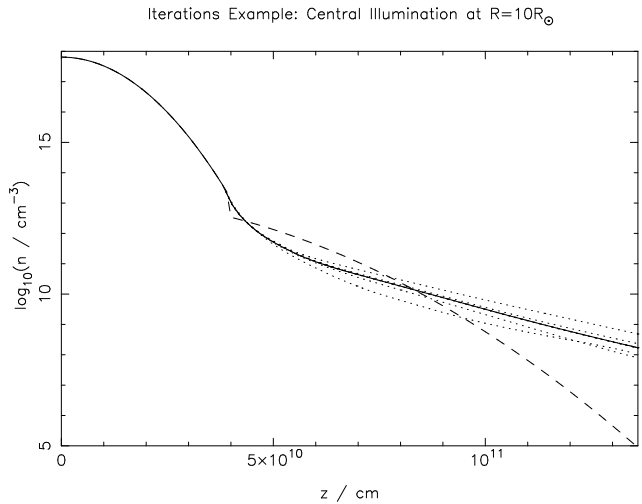}
        \end{turn}
        }
        \caption{Example of iterations towards final density profile.  The inital profile is shown as a dashed line, the subsequent (7) iterations as dotted lines, and the final profile as a solid line.}
        \label{fig:example}
\end{figure}

In this initial model we evaluate the location of the transition point empirically, fixing $z_T$ so that the $T(z)$ profile can be extrapolated smoothly.  (In Section \ref{sec:2d} we define an explicit criterion for the location of the transition point, but this is not necessary at this stage.)  We always define $z_R$ (Equation \ref{eq:den_iter}) to be the point at which the density profile first diverges from that of the isothermal midplane.  An example of the iterative process is shown in Fig.\ref{fig:example}.  Typically this process take 5-10 iterations to converge.

\subsection{Results}
We have first considered uniform illumination of the type expected in the massive star clusters described in Section \ref{sec:bg}.  We evaluated the disc model described above with a uniform X-ray flux $F_X=10^{-2}$erg s$^{-1}$ cm $^{-2}$.  The vertical structure of the disc was evaluated at a number of different radii spanning all of the expected ``static'' region, as well as two models which are outside the gravitational radius.  We evaluated the model at radii of 10R$_{\odot}$ ($=7.0\times 10^{11}$cm), 20R$_{\odot}$ ($=1.4\times 10^{12}$cm), $5.0\times 10^{12}$cm, $1.0\times 10^{13}$cm, $5.0\times 10^{13}$cm, $1.0\times 10^{14}$cm, $2.0\times 10^{14}$cm, $3.0\times 10^{14}$cm, $4.0\times 10^{14}$cm.  At each radius we truncated the disc at the point where the density of the initial profile fell to 1cm$^{-3}$.  

\begin{figure}
        \resizebox{\hsize}{!}{
        \begin{turn}{270}
        \includegraphics{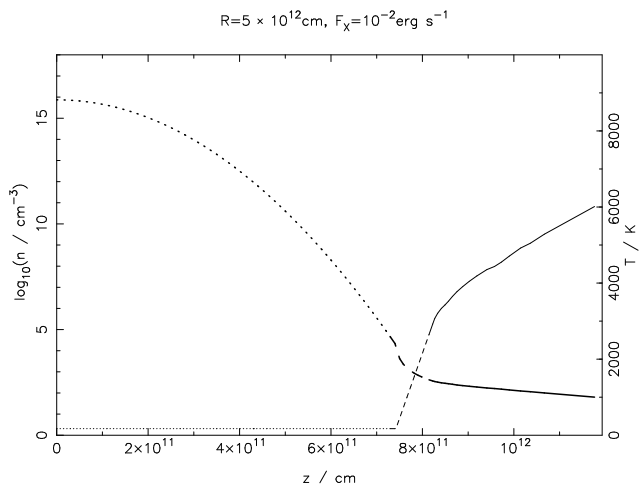}
        \end{turn}
        }
        \caption{Density (heavy line) and temperature (light line) structures evaluated at $R=5\times 10^{12}$cm.  The solid line represents the region calculated explicitly by the model, the dashed line the region where we extrapolate linearly in $T(z)$, and the dotted line the isothermal midplane.}
        \label{fig:r5e12}
\end{figure}
The density and temperature profiles evaluated by the model at $R=5\times 10^{12}$cm are shown in Fig.\ref{fig:r5e12}.  (These are typical of the results at all values of $R$.)  The temperature rises smoothly from the midplane temperature to around 6,000--10,000K.  The gas cannot be heated to temperatures much greater than this without being collisionally ionized, and this only happens in regions where the particle number density is very low ($\lesssim 10^2$cm$^{-3}$).  Thus the region of interest is mostly neutral, with ionization fractions of a few percent at most.  By far the dominant heating process in this region is photoelectronic heating, which accounts for  $>90$\% of the observed heating.  Near to the disc surface direct charge exchange is also significant, typically accounting for a few percent of the total heating.  The contribution of Auger electrons to the total heating is negligible, and the X-rays produced by recombinations (mostly Fe K$\alpha$) are negligible compared to the incident flux.  At the densities of interest here the dominant cooling mechanism is metal line emission.  Only at densities $\gtrsim 10^{10}$cm$^{-3}$ do 2-body cooling processes become significant, so we do not see significant 2-body cooling here.

Due to the large column density of the region heated by X-rays, some three orders of magnitude greater than that for UV photons, the heated region covers a large spatial extent, and in fact expands to a height where the $z \ll R$ approximation (Equation \ref{eq:hydroeq}) is no longer valid.  This is discussed in Section \ref{sec:dis} below.  Also, we note that the outer three models are evaluated outside the scaling radius of 10AU (Equation \ref{eq:surf_den}), in the region where the surface density is falling exponentially.  However, as long as $\Sigma(R)$ is large enough that the unperturbed disc is extremely optically thick at a given radius $R$, the structure of the cold core has little effect on the structure of the heated region.  This condition is satisfied throughout the disc, and so the choice of scaling radius does not affect our results significantly.

Whilst this model is only valid in the inner, static region, a similar structure would be expected in the outer, flow region (although it would obviously be altered somewhat by the flow of heated gas away from the disc).  Thus we can discuss the potential form of mass-loss in a qualitative manner, even if our model cannot account for the process in detail.  As the heating occurs over a much larger column than in the UV case, with a gradual decrease in temperature with decreasing height above the midplane, we would expect the mass-loss profile to differ from the UV case.  In the UV case, where the disc is divided sharply into hot and cold regions at the ionization front, most of the mass is lost at, or very close to, the gravitational radius.  However the smoother temperature profile seen in the our model means that comparable mass loss rates exist at a number of different radii, due to the fact that the escape temperature decreases as $1/R$.  As a result, the point at which the thermal energy of the gas becomes greater than its gravitational energy occurs at a different height above the midplane at each different radius, and so the mass-loss need not be concentrated at the gravitational radius.  Indeed, evaluating $\rho c_s$ at the point where $T(z) = T_{\text {esc}}$ for the outermost three models (which lie at the 8000K gravitational radius and beyond) results in very similar values.  We find mass-loss rates per unit area\footnote{For reference, the UV models of \citet{holl94} derive a mass-loss rate per unit area of $3.88\times 10^{-13}$g cm$^{-2}$ s$^{-1}$ at $R=10^{14}$cm.} of $8.6\times 10^{-17}$g cm$^{-2}$ s$^{-1}$ at $R=2.0\times 10^{14}$cm, $2.0\times 10^{-16}$g cm$^{-2}$ s$^{-1}$ at $R=3.0\times 10^{14}$cm  and $2.5\times 10^{-16}$g cm$^{-2}$ s$^{-1}$ at $R=4.0\times 10^{14}$cm.  However, as mentioned above, this is really over-interpreting our explicitly hydrostatic model, and so these numbers are not considered to be accurate.  

In the very low density regions of the disc, at large $z$, most of the gas is ionized.  Where the gas is mostly ionized, large X-ray fluxes can heat the gas to $\sim 10^6$K, potentially providing some mass-loss at radii as small as 0.1AU.  However the densities in these regions are so low that the maximum mass-loss rates (per unit area) attainable are 2--3 orders of magnitude lower than those at at 10AU, so this effect can safely be neglected.

In principle this seems to be a promising avenue of investigation, with a real potential to influence disc evolution.  However, the problem with these models is the lack of X-ray photons in real systems.  Our models show that significant heating only occurs for values of $F_{\mathrm X} \gtrsim 10^{-4}$erg s$^{-1}$ cm $^{-2}$.  However, \citet{townsley03} observe fluxes of $\simeq 2\times 10^{-7}$erg s$^{-1}$ cm $^{-2}$ in M17 and $\simeq 4 \times 10^{-8}$erg s$^{-1}$ cm $^{-2}$ in the Rosette nebula.  These values are some three orders of magnitude too small to affect the disc structure significantly, and so it seems unlikely that such diffuse bremsstrahlung emission will play a significant role in disc evolution.

\section{Illumination by a central source: 1-D model}\label{sec:central}
We now consider the case of a disc illuminated by X-rays from a central source.  The X-rays emitted from YSOs and TTS have been observed for many years, and a wealth of observational data exist.  As discussed above, the X-rays are subject to rapid variability on small scales, and also occasional ``flaring'' on scales of several orders of magnitude.  If we neglect flaring, however, it is a reasonable approximation to treat the source as having a quasi-steady X-ray luminosity of $10^{28}$--$10^{30}$ erg s$^{-1}$.  Here we assume that the central source has a luminosity of $L_X=10^{30}$ erg s$^{-1}$ and radiates isotropically.  Whilst this may not be valid, the detailed geometry of the X-ray emission is not well enough understood to justify any other selection.  The X-rays are thought to originate from the reconnection of field lines in the upper parts of the magnetosphere, and so, following \citet{gni97}, we place the X-ray source at $R=0$, $z_s=10$R$_{\odot}$.  The validity of this appproximation is discussed in Section \ref{sec:dis} below.

In the case of optical heating of discs by a central source (e.g.~\citealt{as86,kh87,cg97,ddn01}) it is commonly assumed that photons are absorbed at the cylindrical radius where they hit the disc.  Disc heating then occurs through a diffuse radiation field whose net direction, given a disc-like geometry, is approximately vertically downwards.  Thus the flux at each radius is given by:
\begin{equation}\label{eq:x_inten}
F_{\mathrm {X}} = \frac{L_{\mathrm X}}{4 \pi R^2}\frac{z_s}{\sqrt{R^2 + z_s^2}}
\end{equation}
where the first factor in this equation represents simple geometric dilution of the energy and the second term is the vertical component of the incident flux, given a source height $z_s=10$R$_{\odot}$.  The procedure then is to compute the vertical hydrostatic equilibrium structure at each cylindrical radius, subject to this boundary condition on the incident flux.  Although we follow this procedure here as a first guess at the structure of an X-ray irradiated disc, we note that it is less valid in the X-ray case.  The reason for this is that the diffuse radiation field is negligible in the case of X-rays \citep{gni97}, so that heating instead proceeds purely by attenuation (due to absorption) of the incident X-ray beam.  Obviously a self-consistent treatment would involve an iterative solution for the 2-D structure of the disc, subject to hydrostatic equilibrium along each vertical slice and thermal equilibrium along each ray path from the source.  However as a first step we adopt this 1-D approximation.  

The expression for $F_X$ in Equation \ref{eq:x_inten}	 assumes that the X-ray flux is downwardly incident on the disc surface.  Thus it is valid as long as the disc scale height is much less than $z_s$.  If the disc expands to a height greater than $z_s$ this approximation breaks down.  Thus we can use the 1-D model described in Section \ref{sec:uniform} to solve for the disc structure close to the star, subject to this condition on the validity of Equation \ref{eq:x_inten}. 

\subsection{Results}
\begin{figure}
        \resizebox{\hsize}{!}{
        \begin{turn}{270}
        \includegraphics{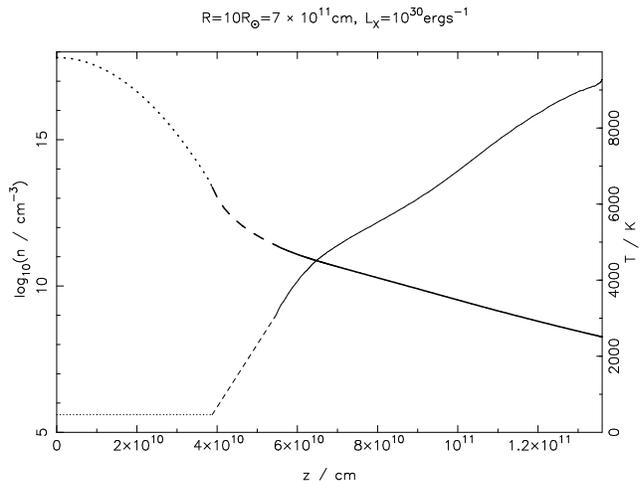}
        \end{turn}
        }
        \caption{As Fig.\ref{fig:r5e12}, but for the central illumination modelevaluated at $R=10$R$_{\odot}$.  The change in gradient of $T(z)$ at $z\simeq6\times 10^{10}$cm is due to the increased rate of 2-body cooling as the density increases.}
        \label{fig:10Rsun}
\end{figure}
\begin{figure}
        \resizebox{\hsize}{!}{
        \begin{turn}{270}
        \includegraphics{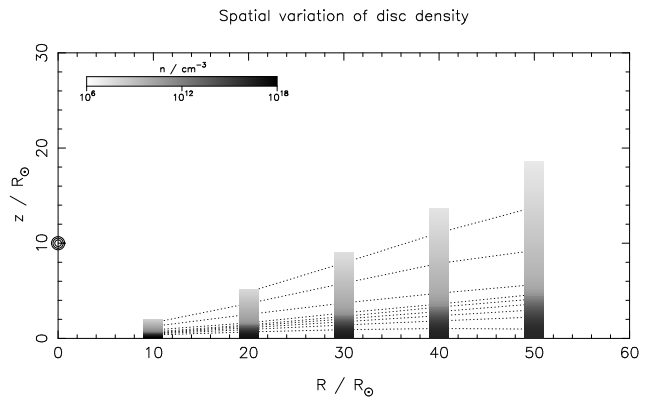}
        \end{turn}
        }
        \caption{Plot of the density distribution produced by the 1-D model.  The greyscale strips are the density profiles evaluated by our model, with dotted contours added as a guide to the eye.  Contours are drawn at $n=10^{16}$, $10^{15}$, $10^{14}$\ldots$10^8$cm$^{-3}$.  The position of the X-ray source is marked by circles.}
        \label{fig:strips}
\end{figure}
We evaluate these models near to the central source, at radii of $10 \mathrm R_{\odot}$, $20 \mathrm R_{\odot}$, $30 \mathrm R_{\odot}$, $40 \mathrm R_{\odot}$ and $50 \mathrm R_{\odot}$.  In general, the structures are very similar to those seen in the uniform radiation field modelled in Section \ref{sec:uniform} above.  The heating rate is much greater, as the local X-ray fluxes so close to the source are much higher ($\sim 10^4$erg s$^{-1}$ cm $^{-2}$), but the heating and cooling processes are very similar.  Due to the high X-ray fluxes we were forced to truncate the initial density profiles at the point where the density falls to $10^4$cm$^{-3}$ to avoid numerical problems in solving for the ionization balance.  The temperature and density profiles resulting from the $R=10 \mathrm R_{\odot}$ model are shown in Fig.\ref{fig:10Rsun}.  The steepening of the $T(z)$ curve as z falls to $\simeq6\times10^{10}$cm is due to the increased significance of 2-body cooling as the density rises.  The overall effect on the structure of the inner disc is shown in Fig.\ref{fig:strips}.

\subsection{Two-component fit}
As seen in Fig.\ref{fig:strips}, at radii $\gtrsim 30$R$_{\odot}$ the disc expands to a height comparable to $z_s$ and so the approximation of downward incidence is no longer valid.  Also, we note that the column density along the line-of-sight from the source to $z_R$ at $R=50$R$_{\odot}$ is $\simeq 10^{26}$cm$^{-2}$.  This corresponds to an optical depth of $\simeq 10^4$ at 1keV, and so it is clear that attenuation of the X-ray flux through the disc will be significant at large radii.  Thus in order to evaluate the global structure of a disc heated by a central X-ray source we must both consider the problem in two dimensions and also account for attenuation of the flux through the disc. 

\begin{figure}
        \resizebox{\hsize}{!}{
        \begin{turn}{270}
        \includegraphics{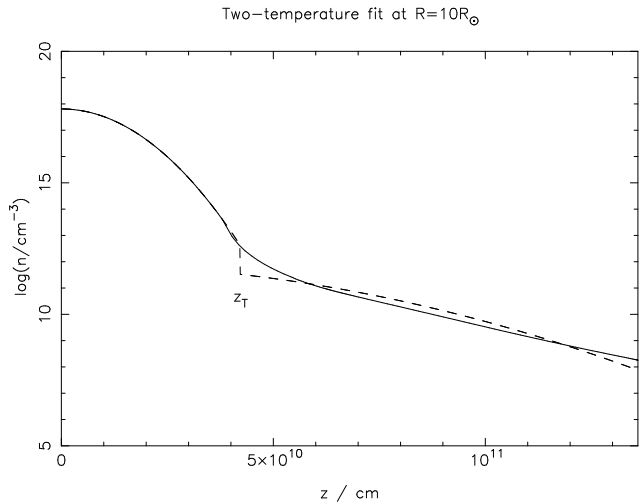}
        \end{turn}
        }
        \caption{Two-temperature fit to the $R=10$R$_{\odot}$ model.  The solid line is the density profile evaluated by the model.  The dashed line is the best-fitting two-temperature profile, with parameters $z_T=4.21\times 10^{10}$cm and $T_{\text {hot}}=6200$K.  The contact discontinuity in the two-temperature density profile at $z_T$ arises because pressure equilibrium is maintained over the temperature discontinuity.}
        \label{fig:bestfit}
\end{figure}

As a first step along the path to a simple 2-D model we note that the vertical structures at these radii can be relatively well-fit by a ``two-component'' profile, parametrized in the same way as the initial profile described in Section \ref{sec:1dmodel} (with free parameters $z_T$ and $T_{\text {hot}}$).  Through a simple least-squares fit we found that a best-fitting temperature of 6200K provided a good fit to the structures at both 10R$_{\odot}$ and 20R$_{\odot}$ (see Fig.\ref{fig:bestfit}).  We will make use of this fact in the next section, where we consider a simple 2-D model.


\section{Illumination by a central source: simple 2-D model}\label{sec:2d}
In order to study the behaviour of the disc at larger radii, in particular nearer to the gravitational radius, we seek to construct a ``toy'' 2-D model of the disc structure.  As seen in Sections \ref{sec:uniform} and \ref{sec:central} above, the vertical density structure of the disc is reasonably well approximated by a two-layer structure, with the cold disc beneath a layer heated by the X-rays.  We have already seen that the vertical structure can be reasonably well-fit by a two-temperature structure, and so an approximate structure of this form provides a good approximation in the parts of the disc which control the photoevaporative mass-loss rate.  The key issue controlling the mass-loss rate is the location of the transition point $z_T$, which marks the bottom of the heated region: to first order, the mass-loss rate per unit area varies as $\rho c_s$, where $\rho$ is the density at $z_T$.  Thus we now seek to solve for the density $n(R,z)$ in a self-consistent manner.

The location of the transition point $z_T$ is determined by the depth to which the X-ray penetrate the disc.  In the optically thin limit, with little or no absorption of the incident flux, the importance of X-ray effects is determined by the parameter
\begin{equation}\label{eq:param}
\xi = \frac{L_X}{n d^2}
\end{equation}
where $L_X$ is the X-ray luminosity, $d$ the distance from the source and $n$ the particle number density at distance $d$.  In essence this parameter is a measure of the photon-to-particle density ratio at any given point (e.g.~\citealt*{tts69,gn01}).  Thus in our model X-ray heating is significant where $\xi$ is greater than some critical value, and where $\xi$ is less than this value it is not.  

Complications arise, however, when optical depth effects become important, as the term $L_X/d^2$ in Equation \ref{eq:param} is effectively a photon flux and so is reduced by extinction effects.  At a single photon energy the optical depth is simply given by $N/N_c$, where $N$ is the column density along the line-of-sight from the source and $N_c$ is the critical column density for which the optical depth is unity.  However as the integrated X-ray photoionization cross-section varies strongly with photon energy $E$, so does $N_c$.  For solar abundances, $N_c$ varies as
\begin{equation}\label{eq:N_c}
N_c = \left(\frac{E}{1{\mathrm {keV}}}\right)^{\alpha} 4.4\times 10^{21} {\mathrm {cm}}^{-2}
\end{equation}
with a best-fitting value of $\alpha=2.485$ \citep{gni97,x_ppiv}.  Thus, with a varying incident continuum ($10^7$K bremsstrahlung) and a varying absorption cross-section we must integrate over photon energy to find the total attenuation.  Using the formulation of \citet{kk83}, updated to use the improved cross-section fits of \citet{gni97}, the attenuation factor $J_h$ is given by the integral:
\begin{equation}\label{eq:Jh}
J_h(\tau,x_0) = \int_{x_0}^{\infty}x^{-\alpha}\exp\left(-x - \tau x^{-\alpha}\right)dx
\end{equation}
where $x_0 = E_0/kT_X$ is the lower cutoff of the spectrum and $\tau$ here is the optical depth to photons with energies at the peak of the continuum (0.7keV).  The first term represents the falling absorption cross-section, and the second is a combination of the exponential factors in the the bremsstrahlung continuum and the extinction.
\citet{gni97} show that $J_h$ can be well-approximated by the expression
\begin{equation}\label{eq:J_fit}
J_h(\tau)=A\tau^{-a}\exp\left(-B\tau^b\right)
\end{equation}
with best fitting parameters of $A=0.800$, $a=0.570$, $B=1.821$ and $b=0.287$ (for solar abundances and keV energies).  Consequently, in the case where attenuation effects are significant the location of $z_T$ is not given by a critical value of $\xi$, but rather a critical value of the modified parameter
\begin{equation}\label{eq:xi_alt}
\xi' = \frac{L_X J_h}{n d^2}.
\end{equation}

Thus our procedure for evaluating the disc struture is as follows.  Firstly we interpolate the inner structure, evaluated in Section \ref{sec:central} above, onto a regular $(R,z)$ grid over the range $R=10$--$20 {\mathrm R}_{\odot}$.  We adopt grid spacings of 2R$_{\odot}=1.39\times10^{11}$cm in $R$ and $1\times10^9$cm in $z$.  In this region the approximation in Equation \ref{eq:x_inten} is valid, and so we use the results of the 1-D model to evaluate the critical value of $\xi'$ at the transition point to be $\xi'_c=4.8\times10^{-8}$erg cm s$^{-1}$.  This uncertainty in this value is approximately 10\%.  The attenuation through the disc to a radius $R_0$ depends only on the structure of the disc at $R<R_0$.  Consequently, if we know the structure in the region $R<R_0$ we can evaluate $J_h$ by integrating the column along the line-of-sight to $(R_0,z)$, and using the fit in Equation \ref{eq:J_fit}.  In this manner we evaluate all values of $\xi'(z)$ at the radius $R_0$.  The height above the midplane which gives $\xi'(z)=\xi'_c$ is adopted as the location of $z_T$ at radius $R_0$, and we then create the resulting vertical structure at $R_0$ using the two-component profile described above (with the best-fitting temperature of $T_{\text {hot}}=6200$K).  We can then move to the next grid cell and repeat the process, and thus iterate outward though the disc to find the global disc structure.

\subsection{Results}
\begin{figure}
        \resizebox{\hsize}{!}{
        \begin{turn}{270}
        \includegraphics{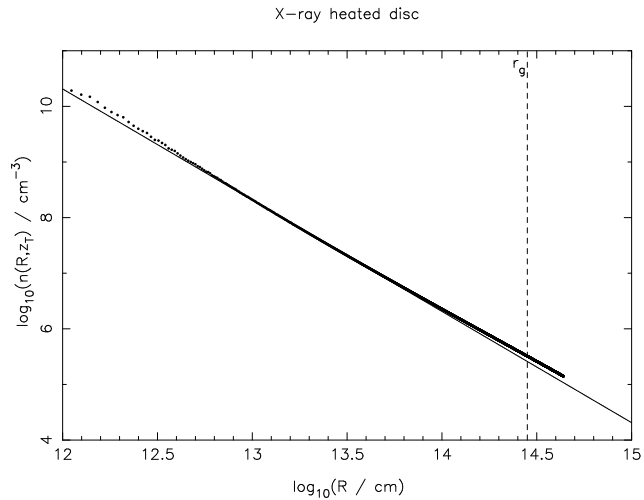}
        \end{turn}
        }
        \caption{Number density at the base of the heated region, , as evaluated by the 2-D model, plotted as a function of radius.  The solid line shows a simple $R^{-2}$ powerlaw.  The only significant deviation from this line is at small radii.  At larger radii the attenuation factor $J_h$ is approximately constant with radius, implying that most of the attenuation occurs close to the star.  The slight deviation from the $R^{-2}$ powerlaw at very large radii is due to the effects of the scaling radius $R_s$ on the density profile, and is not significant.}
        \label{fig:J_h}
\end{figure}
\begin{figure}
        \resizebox{\hsize}{!}{
        \begin{turn}{270}
        \includegraphics{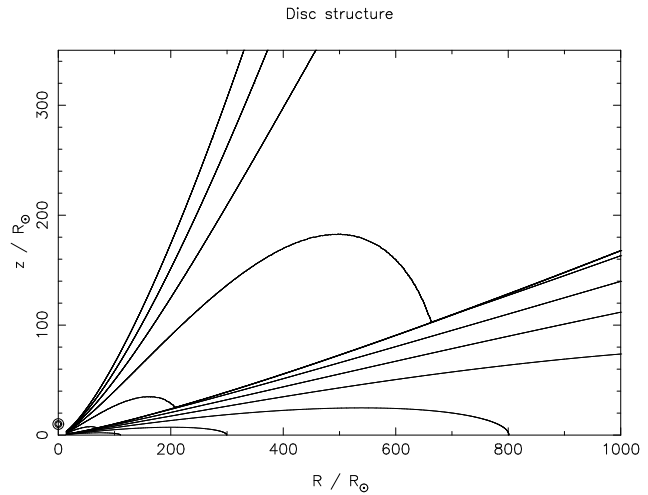}
        \end{turn}
        }
        \caption{Structure of the X-ray heated disc evaluated by the 2-D model.  Density contours are drawn at $n=10^{15}$, $10^{14}$, $10^{13}$\ldots$10^3$cm$^{-3}$.  The location of the X-ray source at $R=0$, $z_s=10$R$_{\odot}$ is marked by circles.}
        \label{fig:global}
\end{figure}
We evaluate the model described above out to a radius of 50AU.  The procedure finds values of $\xi'_c(z_T)$ accurate to within 5\% of the critical value in all cases, and within 1\% for all radii greater than $10^{13}$cm$=0.67$AU.  This translates to an uncertainty of less than 10\% in the value of the density at this point.  Runs at higher spatial resolution ($\times 2$, $\times 5$) over a smaller radial extent do not change the numerical accuracy significantly, and so we consider the procedure to be numerically stable.

The value of $n(R,z_T)$ (i.e.~the number density at the base of the heated region) as a function of radius $R$ is shown if Fig.\ref{fig:J_h}, and the global disc structure is shown in Fig.\ref{fig:global}.  We see from Fig.\ref{fig:J_h} that $n(R,z_T)$ diverges from a $R^{-2}$ power-law at radii $\lesssim100$R$_{\odot}$.  Looking at Equation \ref{eq:xi_alt} we see that this implies that the attenuation factor $J_h$ is constant at radii $\gtrsim100$R$_{\odot}$.  This in turn implies that almost all of the attenuation of the X-ray flux occurs very near to the star.  The slight divergence from the $R^{-2}$ power-law at large radii is a numerical effect due to the value of the scale radius $R_s$ and is not significant.

As before, a first-order estimate of the mass-loss rate per unit area can be made by evaluating $\rho c_s$ at the transition point.  At $2.85\times10^{14}$cm$=19.1$AU, the gravitational radius for 6,200K material, this gives a value of $2.6\times 10^{-13}$g cm$^{-2}$ s$^{-1}$.  Larger disc scale radii $R_s$ can reduce this somewhat, but it remains constant to within a factor of 3 for $R_s<50$AU, and larger scale radii still make little or no difference.  Due to the manner in which we determine $z_T$, neglecting the possibilty of an extended column of ``warm'' material (see Section \ref{sec:1dmodel}), this value of the mass-loss rate is regarded as an upper limit.  As the disc density falls off as $\exp(-z^2/2H^2)$ even small changes in the position of $z_T$ can result in large changes in the value of $n$ at this point, with correspondingly large changes in the mass-loss rate.  Thus we regard this value as an upper limit, with the true mass-loss rate being smaller, possibly significantly so.  The mass-loss rate derived by \citet{holl94} for UV photoevaporation is $3.88\times 10^{-13}$g cm$^{-2}$ s$^{-1}$ at the gravitational radius.  Thus we find that the mass-loss rate of an X-ray driven disc wind is as best comparable to the UV rate, and more likely considerably less significant.

As a consistency check we took the density structure along different lines-of-sight from the source and used these as inputs for the {\sc cloudy} code.  This test indicates that the location of the transition point is accurate to within 10\% at radii less than 1AU.  At larger radii the test confirms that we place the transition point at too low a vertical height.


\section{Discussion}\label{sec:dis}
There are obviously a number of simplifications in these models.  In the 1-D model there is firstly the treatment of the problem as a series of concentric, non-interacting annuli.  This is obviously unphysical and in reality we would expect some radial heat transfer, almost certainly outward through the disc.  However in the limit of vertical heating this radial heat transfer will be negligible by comparison, and so this approximation is sound.

In addition, we point out that our modelling is explicitly hydrostatic, yet attempts to deal with the hydrodynamic behaviour of the disc.  The treatment of the static region, inside the gravitational radius, is valid, as the sound crossing timescale is always much less than the thermal equilibrium timescale.  However outside the gravitational radius the model is clearly not valid, as material will flow away from the disc at a speed comparable to the local sound speed.  Thus the calculations at radii $\gtrsim 2.0\times 10^{14}$cm are suspect, and are considered to be illustrative only.

In the case of a central X-ray source, its location is not well-established.  The X-ray emission is thought to arise from magnetic reconnection events in the stellar magnetosphere, and so should be located in the upper region of the magnetosphere.  Previous studies have used either a point source located at $R=0$, $z=10$R$_{\odot}$ \citep{gni97}, or a ring at $R=5$--$10$R$_{\odot}$, $z=5$--$10$R$_{\odot}$ \citep{ig99,ftb02}.  We adopt the point source for simplicity, and do not investigate the effect of moving the source.  However it is unlikely to affect the results significantly.  We find that the heated disc causes significant attenuation of the X-ray flux for all $R\gtrsim 30$R$_{\odot}$.  Moving the source closer to the disc will likely increase the heating at small radii, but the effect far from the source will be small.

There are several other, more minor problems with the 1-D model which also merit discussion.  One further complication is that the X-ray-heated discs expand significantly in the vertical direction, often well-past the point to which the $z \ll R$ approximation (Equation \ref{eq:hydroeq}) ceases to be valid.  In essence our disc resides in a 1-D harmonic potential, and thus we overestimate the potential at large $z$.  This will lead to the model somewhat underestimating the vertical expansion of the disc.  However the densities of the disc in this region are very small, and so little mass is affected by this simplification.  Also, the manner in which the vertical expansion is evaluated does not explicitly conserve mass.  However the increase in mass due to this simplification only occurs at a level corresponding to around $10^{-8}$ of the total disc mass, and so we neglect its effect.

There is also the issue of grains.  Most of the X-rays are absorbed by heavy elements, which in our model are distibuted evenly throughout the disc (i.e.~with constant grain-to-gas ratio).  However a significant fraction of the heavy elements can be ``locked up'' in grains, and it has been proposed that grains sediment towards the disc midplane (see the review by \citealt*{dust_ppiv} and references within).  This has the knock-on effect of moving the X-ray absorption closer to the disc midplane, thus possibly changing the distribution of heating in the vertical direction.  This issue is not addressed by our model.

Our 2-D model is very simplistic, and is only really a ``toy'' model of the process.  Our criterion for the transition point is correct, in as much as it correctly identifies the point at which the X-rays can heat the disc to temperatures $\gtrsim 3000$K.  However, as discussed in Section \ref{sec:1dmodel} above, the way in which we use the location of this point to create a vertical density structure is somewhat idealised.  By assuming that the temperature continues to fall smoothly we implicitly assume that there is no extended column of ``warm'' material at $\sim 1000$K.  If such a column exists then the point at which the temperature is high enough to drive mass-loss is moved to higher $z$, and therefore somewhat lower density.  Also, the nature of the two-temperature fit is such that we tend to under-estimate the density at $z \gg z_T$.  The approximation of a single temperature in the heated region means that we under-estimate the density where the temperature is much greater than our best-fitting $T_{\text {hot}}=6200$K.  Consequently we under-estimate the extinction along the lin-of-sight to the outer parts of the disc, and therefore over-estimate the depth to which the X-rays penetrate.  A more realistic simulation would incorporate varying temperatures along the different lines-of-sight, but this is beyond the scope of our simple model.  We note however, that both of these simplifications have been made in such a way as to maximise the mass-loss rate.  In addition, we have neglected the possibility that there is any attenuation of the X-ray flux between the source and the disc (apart from that arising in the disc atmosphere), for example by an outflow or an accretion column.  Any reduction in the photon flux reaching the disc would obviously reduce the mass-loss rate still further.  We also note that we have adopted a high X-ray luminosity of $L_X=10^{30}$erg s$^{-1}$.  As can be seen from Equation \ref{eq:xi_alt}, the density at the base of the heated region, and therefore the mass-loss rate, essentially scales as $n \propto L_X$.  Thus a smaller X-ray luminosity, more typical of TTS, will reduce the mass-loss rate still further.  Moreover this linear drop-off is much faster than the square-root dependence on the ionizing flux found in models of UV photoevaporation \citep{holl94}.

Consequently we consider our mass-loss rate to be a robust upper limit to the mass-loss that can be driven by X-ray heating, with the true value being lower, possibly considerably so.  In the even of a significant Lyman continuum flux being present X-ray heating effects will be negligible.  However in the absence of a strong Lyman continuum the X-rays may be able to provide mass-loss at a rate that is significant in disc evolution models.  Further investigation of this problem is still needed, using models that can treat the thermal structure of X-ray irradiated gas over a large dynamic range of temperatures (i.e.~100--10,000K).

Comparisons to previous studies of X-ray/disc interactions are of limited use only, as these studies have had very different goals.  Some studies (e.g.~\citealt{gni97,ig99,ftb02}) have focused on sustaining the very low levels of ionization required to drive the magnetorotational instability.  These studies find that this low level of ionization is provided mostly by hard X-rays (3--10keV), which penetrate much further into the disc than the softer X-rays in which we are interested.  Very few of these hard X-rays are absorbed near to the disc surface, but there are too few hard photons to provide significant heating.  Further, the heated region resulting from our model is in fact optically thin to these hard photons, so the hard X-rays will be absorbed in the cold region close to the midplane.  More recent studies \citep{gn01,gh04} have looked at the observational signatures of X-ray heating.  These arise at lower $z$ and higher density than the region in which we are interested, and so again we are not able to compare to these directly.  Thus our results are consistent with these previous studies, bearing in mind the important caveat that we are interested in very different physical effects.


\section{Conclusions}\label{sec:conc}
We have modelled the effects of X-ray heating on the structure of circumstellar discs, looking in particular at the question of whether X-rays can drive a disc wind.  We first constructed a 1-D model and used it to study the case of uniform X-ray illumination.  Such X-ray emission does exist in some regions of massive star formation, but at a level that is some 4--5 orders of magnitude too low to heat the disc significantly.  We have then looked at heating from a central X-ray source, using parameters typical of T Tauri stars.  Initially we used our 1-D model, but found that it breaks down very close to the central star.  Thus we extended our analysis to consider a simple 2-D model in which X-rays reaching the outer disc are attenuated by passage through the X-ray heated atmosphere of the inner disc.  We find that X-rays are significantly attentuated in this way, with most of the attenuation occuring within $\simeq100$R$_{\odot}$ of the central star.  The critical parameter in determining the mass-loss rate is the location of the transition point $z_T$, which marks the lower boundary at which X-ray heating is significant.  We have therefore constructed a simple 2-D model, solving for $z_T$ in a self-consistent manner.  The model is constructed in such a way as to maximise the mass-loss rate, and most likely over-estimates the mass-loss significantly.  However even our upper limit of $\simeq 10^{-13}$g cm$^{-2}$ s$^{-1}$ is less than that derived for ultraviolet photoevaporation.  A more realistic 2-D simulation, incorporating the thermal response of the gas over a wide range of temperatures, is necessary in order to investigate this process fully.  In the absence of a significant UV flux an X-ray driven disc wind may be significant.  However if a significant UV flux is present, it seems likely that the UV-photoevaporation dominates.  Given the constraints provided by current observational data and the results of this work it seems unlikely that X-ray photoevaporation is a significant factor in the evolution of circumstellar discs around low-mass stars.


\section*{Acknowledgments}
We thank Thierry Montmerle for useful discussions, and also for bringing recent X-ray observations to our attention.  RDA acknowledges the support of a PPARC PhD studentship.  CJC gratefully acknowledges support from the Leverhulme Trust in the form of a Philip Leverhulme Prize.  We thank an anonymous referee for critical comments which greatly improved the paper.


\appendix
\section{Solving the equation of hydrostatic equilibrium}\label{sec:app_a}
The equation of hydrostatic equilibrium (Equation \ref{eq:hydroeq} in Section \ref{sec:basic_model}) governing the vertical structure of an accretion disc is
\begin{equation}\label{eq:hydroeq_ap}
\frac{1}{\rho} \frac{\partial P}{\partial z} = -\frac{GM_*z}{R^3} \quad  \quad (\text{for } z\ll R)
\end{equation}
and the ideal gas law is:
\begin{equation}\label{eq:idealgas}
P=nkT=\frac{\rho k T}{\mu m_{\mathrm H}} \Rightarrow \frac{\rho T}{\mu} = \frac{m_{\mathrm H}}{k} P
\end{equation}
where $\mu$ is the mean molecular weight, $m_{\mathrm H}$ is the mass of the hydrogen atom and $k$ is Boltzmann's constant.  Substutiting $P$ from Equation \ref{eq:idealgas} gives
\begin{equation}
\frac{\partial P}{\partial z} = \frac{k}{m_{\mathrm H}} \left( \frac{\rho}{\mu} \frac{\partial T}{\partial z} + \frac{T}{\mu} \frac{\partial \rho}{\partial z} - \frac{\rho T}{\mu^2} \frac{\partial \mu}{\partial z}\right)
\end{equation}
and so we can substitute this expression into Equation \ref{eq:hydroeq_ap} to find
\begin{equation}
\Rightarrow \frac{1}{T}\frac{\partial T}{\partial z} + \frac{1}{\rho} \frac{\partial \rho}{\partial z} - \frac{1}{\mu} \frac{\partial \mu}{\partial z} = -\frac{m_{\mathrm H}}{k} \frac{GM_*}{R^3} \frac{\mu z}{T}
\end{equation}
Rearranging, we see that
\begin{equation}
\Rightarrow \frac{\partial}{\partial z}\left(\log \frac{\rho T}{\mu}\right) = -\frac{m_{\mathrm H}}{k} \frac{GM_*}{R^3} \frac{\mu z}{T}
\end{equation}
and we can use the ideal gas law again to find that
\begin{equation}
\Rightarrow \frac{\partial}{\partial z} \left(\log P + \log \frac{m_{\mathrm H}}{k}\right) = -\frac{m_{\mathrm H}}{k} \frac{GM_*}{R^3} \frac{\mu z}{T}
\end{equation}
If we now integrate from some reference point $z_R$ to a height $z > z_R$, we find
\begin{equation}
\left[ \log P + \log \frac{m_{\mathrm H}}{k}\right]_{z_R}^z = -\frac{m_{\mathrm H}}{k} \frac{GM_*}{R^3} \int_{z_R}^z \frac{\mu(z') z'}{T(z')} dz'
\end{equation}
and so
\begin{equation}
\Rightarrow P(z) = P_R\exp\left(- \frac{m_{\mathrm H}}{k} \frac{GM_*}{R^3} \int_{z_R}^z \frac{\mu(z') z'}{T(z')} dz'\right)
\end{equation}
where the subscript $_R$ indicates values at $z=z_R$.  Finally we use the ideal gas law again to find the expression for the density profile as a function of the temperature profile and ionization structure.
\begin{equation}\label{eq:den_iter_ap}
n(z)=n_R \frac{T_R}{T(z)}\exp\left(- \frac{m_{\mathrm H}}{k} \frac{GM_*}{R^3} \int_{z_R}^z \frac{\mu(z') z'}{T(z')} dz'\right)
\end{equation}

\label{lastpage}

\begin{thebibliography}{99}
\bibitem[\protect\citeauthoryear{Adams \& Shu}{1986}]{as86} Adams, F.C., Shu, F.H., 1986, ApJ, 308, 836
\bibitem[\protect\citeauthoryear{Alexander, Clarke \& Pringle}{Alexander et al.}{2004}]{columns} Alexander, R.D., Clarke C.J., Pringle, J.E., 2004, MNRAS, 348, 879
\bibitem[\protect\citeauthoryear{Armitage, Clarke \& Tout}{Armitage et al.}{1999}]{act99} Armitage, P.J., Clarke, C.J., Tout, C.A., 1999, MNRAS, 304, 425
\bibitem[\protect\citeauthoryear{Armitage, Clarke \& Palla}{Armitage et al.}{2003}]{acp03} Armitage P.J., Clarke C.J., Palla F., 2003, MNRAS, 342, 1139
\bibitem[\protect\citeauthoryear{Balbus \& Hawley}{1991}]{bh91} Balbus, S.A., Hawley, J.F., 1991, ApJ, 376, 214
\bibitem[\protect\citeauthoryear{Beckwith et al.}{1990}]{beckwith90} Beckwith S.V.W., Sargent A.I., Chini R.S., G\"{u}sten R., 1990, AJ, 99, 924
\bibitem[\protect\citeauthoryear{Beckwith, Henning \& Nakagawa}{Beckwith et al.}{2000}]{dust_ppiv} Beckwith, S.V.W., Henning, T., Nakagawa, Y., 2000, in Mannings, V., Boss. A.P., Russell, S.S., eds, Protostars \& Planets IV.~Univ.~Arizona Press, Tuscon, p.~533
\bibitem[\protect\citeauthoryear{Bell et al.}{1997}]{bell97} Bell, K.R., Cassen, P.M., Klahr, H.H., Henning, Th., 1997, ApJ, 486, 372
\bibitem[\protect\citeauthoryear{Chiang \& Goldreich}{1997}]{cg97} Chiang, E.I., Goldreich, P., 1997, ApJ, 490, 368
\bibitem[\protect\citeauthoryear{Clarke, Gendrin \& Sotomayor}{Clarke et al.}{2001}]{cc01} Clarke C.J., Gendrin A., Sotomayor M., 2001, MNRAS,  328, 485
\bibitem[\protect\citeauthoryear{Dullemond, Dominik \& Natta}{Dullemond et al.}{2001}]{ddn01} Dullemond, C.P., Dominik, C., Natta, A., 2001, ApJ, 560, 957
\bibitem[\protect\citeauthoryear{Duvert et al.}{2000}]{duvert00} Duvert, G., Guilloteau, S., M\'{e}nard, F., Simon, M., Dutrey, A., 2000, A\&A, 355, 165
\bibitem[\protect\citeauthoryear{Eisner \& Carpenter}{2003}]{ec03} Eisner J.A., Carpenter J.M., 2003, ApJ, 598, 1341
\bibitem[\protect\citeauthoryear{Feigelson \& Montmerle}{1999}]{fm99} Feigelson, E.D., Montmerle, T., 1999, ARA\&A, 37, 363
\bibitem[\protect\citeauthoryear{Feigelson et al.}{2003}]{feigelson03} Feigelson, E.D., Gaffney, J.A., Garmire, G., Hillenbrand, L.A., Townsley, L., 2003, ApJ, 584, 911
\bibitem[\protect\citeauthoryear{Ferland}{1996}]{cloudy} Ferland G.J., 1996, {\it A Brief Introduction to {\sc CLOUDY}}, University of Kentucky Department of Physics and Astronomy Internal Report
\bibitem[\protect\citeauthoryear{Font et al.}{2004}]{font04} Font, A.S., McCarthy, I.G., Johnstone, D., Ballantyne, D.R., 2004, ApJ, 607, 890
\bibitem[\protect\citeauthoryear{Fromang, Terquem \& Balbus}{Fromang et al.}{2002}]{ftb02} Fromang, S., Terquem, C., Balbus, S.A., 2002, MNRAS, 329, 18
\bibitem[\protect\citeauthoryear{Gahm et al.}{1979}]{gahm79} Gahm G.F., Fredga K., Liseau R., Dravins D., 1979, A\&A, 73, L4
\bibitem[\protect\citeauthoryear{Glassgold \& Najita}{2001}]{gn01} Glassgold, A.E., Najita, J., 2001, in ASP Conference Series 244, 251
\bibitem[\protect\citeauthoryear{Glassgold, Najita \& Igea}{Glassgold et al.}{1997}]{gni97} Glassgold, A.E., Najita, J., Igea, J., 1997, ApJ, 480, 344 (Erratum: ApJ, 485, 920)
\bibitem[\protect\citeauthoryear{Glassgold, Feigelson \& Montmetle}{Glassgold et al.}{2000}]{x_ppiv} Glassgold, A.E., Feigelson, E.D., Montmerle, T., 2000, in Mannings, V., Boss. A.P., Russell, S.S., eds, Protostars \& Planets IV.~Univ.~Arizona Press, Tuscon, p.~429
\bibitem[\protect\citeauthoryear{Gorti \& Hollenbach}{2004}]{gh04} Gorti, U., Hollenbach, D., 2004, ApJ in press (astro-ph/0405244)
\bibitem[\protect\citeauthoryear{Haisch, Lada \& Lada}{Haisch et al.}{2001}]{haisch01} Haisch K.E., Lada E.A., Lada C.J., 2001, ApJ, 553, L153
\bibitem[\protect\citeauthoryear{Hartmann et al.}{1998}]{hcga98} Hartmann L., Calvet N., Gullbring E., D'Alessio P., 1998, ApJ,  495, 385
\bibitem[\protect\citeauthoryear{Hollenbach et al.}{1994}]{holl94} Hollenbach D., Johnstone D., Lizano S., Shu F., 1994, ApJ,  428, 654
\bibitem[\protect\citeauthoryear{Hollenbach, Yorke \& Johnstone}{Hollenbach et al.}{2000}]{hollppiv} Hollenbach, D.J., Yorke, H.W., Johnstone, D., 2000, in Mannings, V., Boss. A.P., Russell, S.S., eds, Protostars \& Planets IV.~Univ.~Arizona Press, Tuscon, p.~401
\bibitem[\protect\citeauthoryear{Igea \& Glassgold}{1999}]{ig99} Igea, J., Glassgold, A.E., 1999, ApJ, 518, 848
\bibitem[\protect\citeauthoryear{Imanishi, Koyama \& Tsuboi}{Imanishi et al.}{2001}]{ikt01} Imanishi, K., Koyama, K.~\& Tsuboi, Y., 2001, ApJ,  557, 747
\bibitem[\protect\citeauthoryear{Imhoff \& Appenzeller}{1987}]{ia87} Imhoff C.L., Appenzeller I., 1987, in Kondo, Y., ed, Vol.~129, Astroph.~Space Sci.~Lib., Exploring the Universe with the IUE Satellite. Riedel, Dordrecht, p.~295
\bibitem[\protect\citeauthoryear{Kenyon \& Hartmann}{1987}]{kh87} Kenyon, S.J., Hartmann, L., 1987, ApJ,  323, 714
\bibitem[\protect\citeauthoryear{Kenyon \& Hartmann}{1995}]{kh95} Kenyon, S.J., Hartmann, L., 1995, ApJS,  101, 117
\bibitem[\protect\citeauthoryear{Kessel, Yorke \& Richling}{Kessel et al.}{1998}]{kyr98} Kessel, O., Yorke, H.W., Richling, S., 1998, A\&A, 337, 832
\bibitem[\protect\citeauthoryear{Krolik \& Kallman}{1983}]{kk83} Krolik, J.H., Kallman, T.H., 1983, ApJ,  267, 610
\bibitem[\protect\citeauthoryear{Lynden-Bell \& Pringle}{1974}]{lbp74} Lynden-Bell, D., Pringle, J.E., 1974, MNRAS, 168, 603
\bibitem[\protect\citeauthoryear{Mannings \& Sargent}{1997}]{ms97} Mannings V., Sargent A.I., 1997, ApJ, 490, 792
\bibitem[\protect\citeauthoryear{Matsuyama, Johnstone \& Hartmann}{Matsuyama et al.}{2003}]{mjh03} Matsuyama I., Johnstone D., Hartmann L., 2003, ApJ,  582, 893
\bibitem[\protect\citeauthoryear{Neuh\"{a}user}{1997}]{neuhauser97} Neuh\"{a}user, R., 1997, Science, 276, 1363
\bibitem[\protect\citeauthoryear{Pringle}{1981}]{pringle81} Pringle, J.E., 1981, ARA\&A, 19, 137
\bibitem[\protect\citeauthoryear{Richling \& Yorke}{1997}]{ry97} Richling S., Yorke, H.W., 1997, A\&A, 327, 317
\bibitem[\protect\citeauthoryear{Simon \& Prato}{1995}]{sp95} Simon, M., Prato, L., 1995, ApJ, 450, 824
\bibitem[\protect\citeauthoryear{Strom et al.}{1989}]{strom89} Strom K.M., Strom S.E., Edwards S., Cabrit S., Skrutskie M.F., 1989, AJ, 97, 1451
\bibitem[\protect\citeauthoryear{Tarter, Tucker \& Salpeter}{Tarter et al.}{1969}]{tts69} Tarter, C.B,, Tucker, W.H., Salpeter, E.E., 1969, ApJ, 156, 943
\bibitem[\protect\citeauthoryear{Townsley et al.}{2003}]{townsley03} Townsley, L.K., Feigelson, E.D., Montmerle, T., Broos, P.S., Chu, Y-H., Garmire, G.P., 2003, ApJ, 593, 874
\bibitem[\protect\citeauthoryear{Wolk \& Walter}{1996}]{ww96} Wolk, S.J., Walter, F.M., 1996, AJ, 111, 2066
\bibitem[\protect\citeauthoryear{Wyatt, Dent \& Greaves}{Wyatt et al.}{2003}]{wyatt03} Wyatt M.C., Dent W.R.F., Greaves J.S., 2003, MNRAS, 342, 876
\bibitem[\protect\citeauthoryear{Yorke \& Welz}{1996}]{yw96} Yorke, H.W., Welz, A., 1996, A\&A, 315, 555
\end{thebibliography}
\end{document}